\documentclass[preprint,aps,tightenlines,showpacs,nofootinbib]{revtex4}
\usepackage{latexsym}
\usepackage[dvips]{graphicx}
\newcommand{\beq}{\begin{eqnarray}}
\newcommand{\eeq}{\end{eqnarray}}
\newcommand{\eq}{eqnarray}

\newcommand{\al}{{\alpha}}
\newcommand{\be}{{\beta}}
\newcommand{\ci}{\cite}
\newcommand{\ga}{{\gamma}}

\newcommand{\ep}{{\epsilon}}

\newcommand{\de}{{\delta}}
\newcommand{\De}{\Delta}

\newcommand{\m}{{\mu}}
\newcommand{\n}{{\nu}}

\newcommand{\om}{{\omega}}
\newcommand{\Om}{{\Omega}}
\newcommand{\pa}{{\partial}}
\newcommand{\no}{{\nonumber}}
\newcommand{\f}{\frac}
\newcommand{\ra}{\rightarrow}
\newcommand{\lra}{\leftrightarrow}

\newcommand{\tr}{\tilde{r}_+}
\newcommand{\hr}{\hat{r}_+}

\begin{document}

\preprint{arXiv: 0710.1362v4 [hep-th](30 Sept. 2008)}

\title{Smeared BTZ Black Hole from Space Noncommutativity}

\author {Hyeong-Chan Kim$^{a,}$\footnote{E-mail address:
hckim@phya.yonsei.ac.kr}, Mu-In Park$^{b,}$\footnote{E-mail address:
muinpark@gmail.com}, Chaiho Rim$^{b,}$\footnote{E-mail address:
rim@chonbuk.ac.kr}, and Jae Hyung Yee$^{a,}$\footnote{E-mail
address: jhyee@phya.yonsei.ac.kr}}

\affiliation{
$^a$Korea Department of Physics, Yonsei University, Seoul 120-749,
Republic of Korea\\
$^a$Department of Physics and Research Institute of
Physics
and Chemistry, Chonbuk National University, Chonju 561-756, Korea}

\begin{abstract}
We study a novel phenomena of smearing of black hole horizons from
the effect of space noncommutativity. We present an explicit example
in
$AdS_3$ space, using the Chern-Simons formulation of gravity. This
produces a smeared BTZ black hole which goes beyond the classical
spacetime
unexpectedly and there is {\it no} reality problem in our approach
with the gauge group $U(1,1) \times U(1,1)$. The horizons are
smeared, due to a splitting of the Killing horizon and the apparent
horizon, and there is a metric signature change to Euclidean in the
smeared region. The inner boundary of the smeared region acts as a
trapped surface for timelike particles but the outer as a classical
barrier
 for ingoing particles. The lightlike signals
can escape from or reach the smeared region in a {\it finite} time,
which indicates that {\it the black hole is not so dark, even
classically.} In addition, it is remarked that the Hawking
temperature can {\it not} be defined by the regularity in the
Euclidean geometry except in the non-rotating case, and the origin
can be smeared by a {\it new} (apparent) horizon.

\end{abstract}

\pacs{04.70.Dy, 04.60.Kz, 11.10.Nx}
\maketitle

\newpage

\section{Introduction}

It is known that noncommutative field theories \ci{Groe:46,Conn:94}
modify the short distance behaviors of conventional theories. For
example, the noncommutativities can smooth out a singularity of
classical solutions in the conventional theories \ci{Doug:01}.
However, similar modifications in gravity theories are not studied
much,
though there are several formulations of noncommutative gravities
and speculations. For example, one can generally expect a
``smearing'' of the horizons, which are sharply defined in the
conventional spacetime. From the commutation relations
\begin{\eq}
\label{t:cart} [ x^i, x^j ]=i \theta^{ij},
\end{\eq}
where $\theta^{ij}$
is an antisymmetric constant of dimension $length^2$,
the precise location of a horizon
is limited by the uncertainty relations $\De x^i \De x^j \ge
|\theta^{ij}|/2$. But, there has been no {\it explicit}
demonstration of existence of the smearing region with unexpected
space-time structure.

In this paper, we investigate the modifications of the BTZ black
hole in three-dimensional anti-de Sitter ($AdS_3$) space with a
polar form of the commutation relation
\begin{\eq}
\label{rphi}
 [ r, \phi ]=i \hat{\theta}
\end{\eq}
for a spherically symmetric case ($ \theta^{r \phi}\equiv
\hat{\theta}$)\footnote{This commutation relations differ from the
Cartesian ones with a constant $\theta^{ij}$ \ci{Pinz:06} since it
corresponds to a non-constant $\theta^{ij}=r \hat{\theta} \ep^{ij}$
in (\ref{t:cart}). But, the Moyal product \ci{Groe:46} can be still
consistently defined in the polar coordinate with a constant
$\theta^{r \phi}$. And we expect no big difference in the {\it
qualitative} physics near the horizon since we are considering
physics at $r \neq 0$.}, using the Chern-Simons formulation of
gravity. This is the first explicit example of the novel phenomena
of the horizon smearing which goes beyond the classical spacetime
unexpectedly, and {\it without} the reality problem by considering
the gauge group $U(1,1) \times U(1,1)$. It is found that the event
horizon is {\it smeared} by the splitting of the Killing horizon and
the apparent horizon with the thickness of the order $\hat{\theta}$
and the smeared horizon acts as a classical barrier for particles.
The time duration of light signals escaping from and reaching the
smeared
horizon region from outside is demonstrated to be finite. The
physical effect
of the metric signature change inside the smeared
region
is considered. In addition, it is remarked that the Hawking
temperature from the regularity in the Euclidean geometry is {\it
not}
applicable, and the naked conical singularity at the origin in
$AdS_3$ or negative mass solutions, generally, is smeared due to the
presence of a ``new'' apparent horizon near the origin.

\section{Three-dimensional noncommutative gravity from
Chern-Simons formulation} \label{csformulation}

It is well known that, in three-dimensional space-time, conventional
gravity theory can be formulated as a Chern-Simons theory
\ci{Achu:86}. This provides a novel way to define the noncommutative
gravity theory \ci{Bana:00}.
An explicit solution in the noncommutative gravity can be obtained
from the corresponding known solution in the commutative gravity,
via the Seiberg-Witten map \ci{Gran:00}.

The (2+1)-dimensional noncommutative gravity with the negative
cosmological constant $\Lambda=-1/l^2$ is defined by the
$U(1,1) \times U(1,1)$ Chern-Simons theory, up to surface terms,
\begin{\eq}
\label{NCg}
 I_g [\hat{\cal A}] =\f{l}{16 \pi G} \int 
 Tr \left(\hat{\cal
A}^{+} \star d \hat{\cal A}^{+} +\f{2}{3} \hat{\cal A}^{+} \star
\hat{\cal A}^{+} \star \hat{\cal A}^{+} \right)- \left(\hat{\cal
A}^{+} \lra \hat{\cal A}^{-} \right).
\end{\eq}
(Here, the wedge symbol has been omitted.) The connections are given
by \footnote{We take the
$SU(1,1)$ bases $t_0=\sigma_2/2, t_1=i \sigma_3/2, t_2=\sigma_1/2$
such that $[t_a, t_b]=-\epsilon_{ab}^c t_c, Tr (t_a t_b) =(1/2)
\tilde{\eta}_{ab}$ with $\epsilon_{012}=1$ and
$\tilde{\eta}_{ab}=diag(+1,-1,+1).$}
\begin{\eq}
\hat{\cal A}^{\pm}=\left(\hat{\omega}^a
\pm {\hat{e}^a
}/{l} \right) t_a + i \hat{c}
^{\pm} {\bf 1}
\end{\eq}
with the triads and the
$SU(1,1)$ spin connections $\hat{e}^a=\hat{e}^a_{\mu}
dx^{\mu},~\hat{\omega}^a_{\mu}=(1/2) \epsilon^{abc}
\hat{\omega}_{\mu bc}dx^{\m}~(a=0,1,2)$, respectively, and the
$U(1)$ connections $c^{\pm}$ which make the group closed with
respects to the Moyal $\star$ product \ci{Groe:46,Bana:00}
\begin{\eq}
\label{star} \star = \mbox{exp} \left[{\frac{i}{2} \hat{\theta}
(\overleftarrow{\partial_r}
\overrightarrow{\partial_{\phi}}-\overleftarrow{\partial_{\phi}}
\overrightarrow{\partial_r} )}\right].
\end{\eq}
Here, it is important to note that, in the commutative limit, the
theory does not depend on the metric, i.e., the choice of the
coordinates \ci{Witt:89}\footnote{The metric dependence in the
Cartesian measure of integration, $d^3 x$, under the general
coordinate transformations, $d^3 x'=d^3 x /\sqrt{g}$, is canceled by
that of the Levi-Civita tensor density  $\ep^{\al \be \ga}=\sqrt{g}
\pa_{\m} x'^{\al} \pa_{\n} x'^{\be} \pa_{\rho} x'^{\ga} \ep^{\m \n
\rho}$ in the Chern-Simons $3$-form
$d^3 x  \epsilon^{\m \n \rho} Tr[{A}_{\m} \pa_{\n} { A}_{\rho}
+({2}/{3}) { A}_{\m} { A}_{\n} { A}_{\rho} ]$.
}. In the polar coordinates
$\m=(t, r, \phi)$, for example, the (commutative) action takes the
form \ci{Cous:95}:
$
 I_g [{A}] =({l}/{16 \pi G}) \int dt dr d \phi
~ Tr (- {A}^{+}_{r} \partial_t {A}^{+}_{\phi}
 + {A}^{+}_{\phi}   \partial_t {A}^{+}_{r} +2 {A}^{+}_{t}
 { F}^{+}_{r \phi} )-({A}^{+}_{\m} \lra {A}^{-}_{\m} ).
$
We consider the Moyal $\star$ deformation of this polar action
as follows:
\begin{\eq}
\label{NCg}
 I_g [\hat{\cal A}] =\f{l}{16 \pi G} \int dt dr d \phi
~ Tr \left(- \hat{\cal A}^{+}_{r} \star \partial_t \hat{\cal
A}^{+}_{\phi}
 + \hat{\cal
A}^{+}_{\phi}  \star \partial_t \hat{\cal A}^{+}_{r} +2 \hat{\cal
A}^{+}_{t} \star  \hat{\cal F}^{+}_{r \phi} \right)- \left(\hat{\cal
A}^{+}_{\m} \lra \hat{\cal A}^{-}_{\m} \right).
\end{\eq}
(We define
$\hat{\cal F}^{\pm}_{\mu \nu}=\pa_{\m} \hat{\cal
A}^{\pm}_{\n}+\hat{\cal A}^{\pm}_{\m} \star \hat{\cal
A}^{\pm}_{\n}-(\m \lra \n)$, in a covariant way.)
We note that the noncommutative action has the measure of integration 
$dt dr d \phi$ and so the Moyal product (\ref{star}) is well
defined,
as
in the usual Cartesian coordinates \ci{Gran:00}.
There are some differences in the global properties of the
coordinates, due to the range of the coordinates $(0, \infty) \times
(0, 2 \pi)$ and so there will be some appropriate boundary
conditions on the
allowed functions (see footnote 4, for example).
However all the standard ``local''
(i.e., ignoring boundary conditions) formula in Cartesian
coordinates, like equations of motion and the Seiberg-Witten map,
will work also here \cite{Bana:00}.

The equations of motion are given by
\begin{\eq}
\label{NCF=0} d \hat{\cal A}^{\pm} +\hat{\cal A}^{\pm} \star
\hat{\cal A}^{\pm} =0
\end{\eq}
which are not easy to solve directly. But using the Seiberg-Witten
map \ci{Seib:99}, which transforms the noncommutative Chern-Simons
action into the commutative one, or vice versa\footnote{In this
mapping, there are several boundary terms which do not vanish for an
arbitrary choice of $\theta^{ij}$, generally. But, these terms
vanish for our choice (\ref{rphi}) and the commutative solutions
${\cal A}^{\pm}$ which decrease rapidly for large $r$.}, without
additional action \ci{Gran:00}, any solution ${\cal A}^{\pm}$ of the
commutative equations $ d{\cal A}^{\pm} +{\cal A}^{\pm} {\cal
A}^{\pm} =0 $ can be mapped into the corresponding $\hat{\cal
A}^{\pm}$ of the noncommutative equations (\ref{NCF=0})
\ci{Bono:00}:
\begin{\eq}
\label{SW}
 \hat{\cal A}^{\pm}_{\m} (\theta)={\cal A}^{\pm}_{\m}+
(i/4) \theta^{\al \be} [ {\cal A}^{\pm}_{\al}, \pa_\be {\cal
A}^{\pm}_{\m} + {\cal F}^{\pm}_{\be \m} ]_{+} + {\cal O} (\theta^2).
\end{\eq}
From the obtained solution $\hat{\cal A}^{\pm}$, one can compute
$\hat{\om}$ and $\hat{e}$ which describe the noncommutative gravity.

\section{BTZ black hole solution with U(1)
fluxes in $U(1,1) \times U(1,1)$
Chern-Simons gravity}
\label{BTZblackhole}

In the commutative limit, the equations of motion reduce to the  two
sets of {\it decoupled} equations
\begin{\eq}
\label{e_o}
&& d \om +\om \om -(1/l^2) e e=0,~d e + \om e + e \om =0, \\
\label{dc} && d c^{\pm}=0.
\end{\eq}
We
generalize these to the case with the non-vanishing $U(1)$ fluxes
 $d c^{\pm} =f^{\pm}$ by adding an additional term $-2 \int
Tr({f} {\cal{A}})$ to the Chern-Simons gravity action. This
modifications do not change the conventional gravity equations
(\ref{e_o}), since these are decoupled from the $U(1)$-parts
(\ref{dc}).

In the noncommutative case,  we have the additional noncommutative
action $-2 \int Tr(\hat{f} \star \hat{\cal{A}})$ \ci{Poly:00}. One
can confirm that this term is invariant under the Seiberg-Witten
map,
for the appropriate fluxes $\hat{f}$ which decrease rapidly for
large $r$ with our choice (\ref{rphi}). In consequence, the solution
$\hat{\cal{A}}$ of the generalized theory can be simply obtained by
the same mapping (\ref{SW}) from the corresponding commutative
solution $\cal{A}$. The $U(1)$ fields are not decoupled anymore and
have non-trivial effects on the commutative gravity solutions.

To proceed, we consider the Aharonov-Bohm type $U(1)$ potentials
\begin{\eq}
c^{\pm}=\Phi^{\pm} d \phi
\end{\eq}
which give the fluxes inside the horizons, $f^{\pm}=2 \pi \Phi^{\pm}
\de ^2 ({ \bf x}) dr d \phi$.
For the (commutative) gravity solution, we consider the BTZ black
hole
given
by \ci{Bana:92}
\begin{\eq}
\label{BTZ}
 ds^2=-N^2 dt^2 +N^{-2} dr^2 +r^2 (d \phi +N^{\phi} dt)^2
\end{\eq}
with
$
\label{N}
 N^2={(r^2-r_+^2)(r^2-r_-^2)}/{l^2 r^2},~N^{\phi}=-r_+
{r_-}/{l r^2}. $
Here, $r_+$ and $r_-$ denote the outer and inner horizons,
respectively. The $SU(1,1) \times SU(1,1)$
1-form gauge connections ${\bf A}^{\pm} =(\om^a \pm e^a/l )t_a$ are
given by \ci{Carl:95}
\begin{eqnarray}
{\bf A}^{\pm}= \f{1}{2} \left(
\begin{array}{cc}
  \pm  i ( l/\n) d \m & z_{\pm} (\n \mp i \m)  (dt/l \mp d \phi) \\
 z_{\pm} (\n \pm i \m) (dt/l \mp d \phi)  & \mp i ( l/\n) d \m
\end{array}\right),
\end{eqnarray}
where $\n^2 (r)=(r^2-r_-^2)/(r_+^2 -r_-^2), \m^2=\n^2-1$, and
$z_{\pm}=(r_+ \pm r_-)/l$.

For the noncommutativity relations (\ref{rphi}) with $\hat{\theta}$
rescaled as  $\hat{\theta}=l \theta$ and others$=0$, the solutions
$\hat{\cal{A}}^{\pm}$ in the noncommutative gravity (\ref{NCg}) are
obtained, via the Seiberg-Witten map,
\begin{\eq}
\label{BTZ:SW} &&\hat{\cal{A}}_t^{\pm}=i ( l \theta/4) Tr(
A_{\phi}^{\pm} \pa_r A_t^{\pm} )
{\bf 1} +[1+(l \theta/2) c_{\phi}^{\pm} \pa_r ]A_t^{\pm},  \no \\
&&\hat{\cal{A}}_r^{\pm}=i ( l \theta/4) Tr( A_{\phi}^{\pm} \pa_r
A_r^{\pm} ){\bf 1} +[1+(l \theta/2) \{(\pa_r c_{\phi}^{\pm})
+c_{\phi}^{\pm} \pa_r\} ]A_r^{\pm}, \no \\
&&\hat{\cal{A}}_\phi^{\pm}=[1+(l \theta/2) (\pa_r c_{\phi}^{\pm}) ]
{\cal A}_{\phi}^{\pm},
\end{\eq}
(note that ${\cal A}_{t,r}^{\pm}=A_{t,r}^{\pm}, {\cal
A}_{\phi}^{\pm}=c^{\pm}_{\phi}+A_{\phi}^{\pm}$), neglecting the
higher order terms of ${\cal O} (\theta^2)$. The metric of the
noncommutative gravity is  defined\footnote{Here, the metric has no
``reality problem'' since the equality $\hat{e}_{\m} \star
\hat{e}_{\n}=\hat{e}_{\m} \hat{e}_{\n}$ holds in our metric
(\ref{BTZ}) due to the commutation relations (\ref{rphi}). This is a
unique feature of our approach and in sharply contrast to the
previous approaches \ci{Cham:01} where the metric becomes {\it
complex} generally and some truncations are needed to get a real
metric.} as $d \hat{s}^2 =\eta_{ab} (\hat{e}^a_{\m} \star
\hat{e}^b_{\n})dx^{\m} dx^{\n}$ with $\eta_{ab}=diag(-1,+1,+1)$ and
given by
\begin{\eq}
\label{NCBTZ} d \hat{s}^2=-f^2 dt^2 +\hat{N}^{-2} dr^2 +r^2 (d \phi
+N^{\phi} dt)^2 +  {\cal O} (\theta^2),
\end{\eq}
where
\begin{\eq}
\label{N:NC}
 &&\hat{N}^2=\f{r^2}{l^2}  +\theta c_{\phi} \f{r}{l}
-\f{(r_+^2+r_-^2)}{l^2} +\f{r_+^2 r_-^2}{l^2 r^2}-\theta c_{\phi}
\f{r_+^2 r_-^2}{l r^3}, \\
&&f^2=\hat{N}^2+\theta c_{\phi} \f{r_+^2 r_-^2}{l r^3}.
\end{\eq}
Here, we consider $c_{\phi}^{+}=c_{\phi}^{-}$ for simplicity and
omit the singular term at the origin $-2 \pi l \theta N^2 \Phi \de^2
({\bf x})$ in $\hat{N}^2$ since we are considering physics at $r\neq
0$. Note that, in this case, the noncommutative spacetime satisfies
the same {\it gravity} equations of motion as in the commutative
limit, i.e., $
d \hat{\om} +\hat{\om} \hat{ \om} -(1/l^2) \hat{e} \hat{ e}=0, ~d
\hat{e} + \hat{\om}{\hat e} + \hat{e} \hat{\om} =0,
$ with {\it no} Moyal products, from (\ref{NCF=0}) and the
triviality of bi-products in our solution (\ref{BTZ:SW}), i.e.,
$\hat{\omega} \star \hat{\omega}=\hat{\omega} \hat{\omega}$,
$\hat{\omega} \star \hat{e}=\hat{\omega} \hat{e}$, etc. In other
words, the (first-order) noncommutative solution (\ref{BTZ:SW}) has
the same noncommutative curvature $\hat{R} \equiv d \hat{\om}
+\hat{\om} \hat{ \om} = (1/l^2) \hat{e} \hat{ e}$ and (zero) torsion
$\hat{T}\equiv d \hat{e} + \hat{\om}{\hat e} + \hat{e} \hat{\om}
=0$, as in the conventional BTZ black hole spacetime, outside the
point flux \cite{Carl:95}.\footnote{This may be compared with a BTZ
solution in the presence of higher derivative terms, like the
gravitational Chern-Simons action term. There the BTZ solution
satisfies, trivially, the Einstein's equations of motion, with no
higher derivative contributions. However, this does not mean a
triviality of the solution since its physical parameters, like the
ADM mass and angular momentum are significantly deformed from the
conventional ones \cite{Park:07}.
}

\section{Properties of the smeared black hole}
\label{properties}

The noncommutative black hole solution has several remarkable
properties which go beyond the classical geometry,
unexpectedly.\\

1. {\it There is a splitting of the apparent horizon and the Killing
horizon}: The apparent horizon is defined as a null hypersurface
$g^{\m \n}(\pa_{\m} r)(\pa_\n r)=\hat{N}^2=0$, whereas the Killing
horizon as the surface where the norm of the Killing vector
$\chi=\pa_t +\Om_H \pa_{\phi}$ vanishes, i.e., $\chi^2=g_{tt}-(g_{t
\phi})^2/g_{\phi \phi}=-f^2=0$ with the angular velocity of the
horizon $\Om_H=-(g_{t \phi}/g_{\phi \phi})|_{H}$.

We first note that $\hat{N}^2=[x^5 +\theta c_{\phi} x^4
-(x_+^2+x_-^2)x^3 +(x_+^2 x_-^2)x-\theta c_{\phi} x_+^2 x_-^2]/x^3
=0~ (x  \neq 0,~ x \equiv r/l)$ has the outer/inner (apparent)
horizons at
\begin{\eq}
\label{AH}
 \hat{r}_{\pm}=r_{\pm} -l \theta c_{\phi}/2.
\end{\eq}
The apparent horizons are {\it equally} shifted by the small amount
$-l \theta c_{\phi}/2$. On the other hand, the outer/inner Killing
horizons can be obtained from
 $f^2=[x^4 +\theta c_{\phi} x^3 -(x_+^2+x_-^2)x^2
+x_+^2 x_-^2]/x^2 =0$,
\begin{\eq}
\label{KH}
 \tilde{r}_{\pm}=r_{\pm} -( l \theta
c_{\phi}/2)(1-r_{\mp}^2/r_{\pm}^2)^{-1}.
\end{\eq}
This shows that Killing horizons are
{\it not}
equally shifted
so that the apparent and Killing horizons do not coincide each other
in general, except in
the
{\it non-rotating}
BTZ ($r_-=0$) \footnote{However, the
inner horizons $
\hat{r}_{-}, \tilde{r}_{-}$ are absent in this case, though not
manifest in (\ref{AH}) -(\ref{KH}). }. It is noted that
the solution (\ref{KH}) of the Killing horizons are not valid
near the extremal commutative black holes with $r_+=r_-$
where the higher order corrections are needed in contrast to the apparent horizons (\ref{AH}).

2. {\it The event horizon becomes ``smeared'', due to the splitting
of the Killing and apparent horizons}: To see this, let us consider
the metric
\begin{\eq}
\label{metric:co-ro} d \hat{s}^2=-f^2 dt^2 +\hat{N}^{-2} dr^2 +r^2
(d \tilde{\phi} +\tilde{N}^{\phi} dt)^2,
\end{\eq}
such as $\tilde{N}^{\phi}=0$ either at the Killing horizons
$\tilde{r}_+$ or the apparent horizon $\hat{r}_+$, for an
appropriate choice of the {\it co-rotating} frame. Then, the radial
null geodesics are given by
\begin{\eq}
\label{geod}
 dr/dt=\pm \sqrt{f^2 \hat{N}^2 },
\end{\eq}
with the upper (lower) sign corresponding to outgoing (ingoing)
geodesics.

Near the Killing horizon $\tilde{r}_+$, the radial null geodesics
for $r>\tilde{r}_{+}$
 are given by
\begin{\eq}
\label{geod:tr}
 dr/dt=\pm \sqrt{2 \tilde{\kappa} (- \theta c_{\phi}) r_+^2 r_-^2/l
\tilde{r}_+^3 } \sqrt{r- \tilde{r}_+},
\end{\eq}
where $\tilde{\kappa}$ corresponds to the surface gravity in the
usual context ($f^2 \approx 2 \tilde{\kappa} (r-\tilde{r}_{+})$):
\begin{\eq}
\tilde{\kappa}&=&(1/2) (\pa f^2/\pa r) |_{\tr} \no \\
&=& (\tr^4-r_+^2 r_-^2)/(l^2 \tr^3) +\theta c_{\phi}/2l\,.
\end{\eq}
For the non-negative $\tilde{\kappa}$, which is always the case when
$\theta$ is not so large, the outgoing as well as the ingoing
geodesics are ``classically'' allowed only if
the velocity (\ref{geod:tr}) is real,
i.e., $\theta c_{\phi} <0$,  or  $r_+ < \hr < \tr$.

Even though the light cones close up as we approach the
horizon $\tr$, which signals usually that the time coordinate $t$ is
badly defined near the horizon,
it is remarkable that the light signals can escape from
and reach the horizon $\tr$ in a {\it finite} time
\begin{\eq}
\label{time:tr}
 \tilde{t} \approx
[\tilde{\kappa} (- \theta c_{\phi}) r_+^2 r_-^2/2 l
\tilde{r}_+^3]^{-1/2} \sqrt{r -\tr}.
\end{\eq}
This is in contrast to the conventional commutative case\footnote{Note
that this behavior can {\it not} be directly obtained by setting
$\theta c_{\phi}=0$ in (\ref{time:tr}) since we must consider
$(r-r_+)$-term again which now dominates $\theta c_{\phi}$-term, in
that case.},
where one needs an infinite time $t \sim
ln(r-r_+)$ to escape from the horizon $r_+$ though a finite
``proper'' time to reach the horizon \ci{Pari:00}. It seems that the
singular behavior of the time coordinate $t$ near the horizon is
``moderated'' by the noncommutativity effect. Thus, the horizon is
smeared and {\it not so ``dark'', even classically !}

The same is true when $\theta c_{\phi}>0$,
or
 $ \tr < \hr < r_+$. In this case,
the radial null
geodesics
 for $r > \hr$
near the apparent horizon $\hr$ is  given by
\begin{\eq}
\label{geod:hr} dr/dt=\pm \sqrt{2 \hat{\kappa} (\theta c_{\phi})
r_+^2 r_-^2/l \hat{r}_+^3 } \sqrt{r- \hat{r}_+},
\end{\eq}
where $\hat{N}^2 \approx 2 \hat{\kappa} (r-\hat{r}_{+})$ and $ f^2
\approx ( 2 \hat{\kappa} -3 \theta c_{\phi} r_+^2 r_-^2/l \hr^4)
(r-\hr) +\theta c_{\phi} r_+^2 r_-^2/l \hr^3$ are used with
\begin{\eq}
\hat{\kappa}&=&(1/2) (\pa \hat{N}^2/\pa r) |_{\hr} \no \\
&=& (\hr^4-r_+^2 r_-^2)/(l^2 \hr^3) +\theta c_{\phi}/2l +3 \theta
c_{\phi}r_+^2 r_-^2/(2l \hr^4).
\end{\eq}
The
geodesics show
that the  escaping time from (or approaching time to) the horizon
$\hr$ is finite
\begin{\eq}
\label{time:hr} \hat{t} \approx
 [\hat{\kappa} ( \theta c_{\phi}) r_+^2 r_-^2/2 l
\hat{r}_+^3]^{-1/2} \sqrt{r -\hr}~.
\end{\eq}

3. {\it The smeared horizon region behaves as a barrier for
particles and waves}: This comes from the fact that $f^2 \hat{N}^2
<0$ in the region between $\tr$ and $\hr$ and one has the {\it
imaginary} radial velocities for the (radial) null geodesics
(\ref{geod})\footnote{This can be also observed in (\ref{geod:tr})
and (\ref{geod:hr}).}. This is the consequence of the fact that
there is a ``signature change''\footnote{This seems to reflect the
{\it quantum gravity} nature of the noncommutative geometry, as is
in the beginning of the Universe \ci{Hart:83}.
} to Euclidean
 $(+++)$, in the smeared region
i.e., there is ``no-time'' and there are no light cones for the
radial motions when $\theta c_{\phi}<0$. In this sense, the outer
(horizon) boundary becomes {\it hard} to penetrate for particles,
compared to the conventional event horizon. Nevertheless, light wave
\footnote{Particles may tunnel {\it quantum mechanically}, by their
wave nature also.} may tunnel the smeared horizon region when its
wavelength is greater than the thickness of the region.
 (Similar thing happens when $\theta c_{\phi}>0$. In this
case, we have (pseudo) Euclidean geometry with the signature
$(--+)$, i.e., there are ``two-times'' in the smeared region, and
time and radial coordinates change the role.)

However, particles or waves inside the inner boundary can not escape
from the black hole since the light cone structures in that region
are the same as in the interior region of the event horizon of the
commutative case. So, the inner boundary of the smeared region is
the trapped surface. The usual Hawking radiation would be generated
near the inner boundary since one of the pair-created particles can
be trapped. On the other hand, the pair-created particles near the
outer boundary always recombine, due to the absence of the trapping.
Thus, our result seems to favor the tunneling picture of the Hawking
radiation by Parikh and Wilczek \ci{Pari:00}. Further studies are
needed in this direction.

4. {\it The Hawking temperature defined by the periodicity in the
Euclidean time is not applicable:} To see this, let us consider the
metric (\ref{metric:co-ro}) near the Killing horizon $\tr$.
Following the usual approach, we put the Euclidean time $\tau=-it$
and we get
\begin{\eq}
d \hat{s}^2 \approx \tilde{\kappa}^2 \eta^2 d \tau^2 +\left[
\left(1+\f{3 \theta c_{\phi} r_-^2}{2 l r_+^2 \tilde{\kappa}}
\right) -\f{\theta c_{\phi} r_-^2}{l r_+ \tilde{\kappa}^2
\eta^2}\right]^{-1} d \eta^2 + (\tr +\tilde{\kappa} \eta^2/2 )^2 (d
\tilde{\phi})^2,
\end{\eq}
where  $\eta=\sqrt{2 (r- \tr)}$. From the regularity of the
$d\tau^2$-part, one obtains the periodicity $\be=2 \pi
/\tilde{\kappa}$ which would give the conventional Hawking
temperature $T_H=\tilde{\kappa}/2 \pi$ \ci{Hart:76}. For the
non-rotating case ($r_{-}=0$), the system is quite normal and the
Killing and apparent horizons coincide, except the shift of the
horizon in (\ref{AH}). In this case, the Hawking temperature can be
defined as usual with a constant shift\footnote{If one considers the
Bekenstein-Hawking entropy \ci{Beke:73}, $\hat{S}_{BH}=2 \pi \hr$
with respect to the horizon $\hr$, one would have the first law of
thermodynamics with the black hole mass $\hat{M}=(\hr^2 +l \theta
c_{\phi} \hr)/2 l^2$, up to some additive constant terms. But, it is
not clear whether the Bekenstein's area law
is satisfied even in the noncommutative geometry, which can be
regarded as the higher derivative gravities \ci{Park:07}. }:
$ \hat{T}_H =(\hr +l \theta c_{\phi}/2)/(2 \pi l^2). $

However, the (near horizon) geometry
spoils the regularity in general since, as $\eta \ra 0$, the radial
coordinate $\eta$ becomes time-like and the Euclidean procedure
itself becomes invalid. Thus, the conventional way of defining the
temperature is not valid anymore and this might be a general
phenomena for the noncommutative geometry with the smeared horizons.
Somehow we suspect that $T_H=\tilde{\kappa}/2 \pi$, which converges
to the usual Hawking temperature $T_H={\kappa_+}/2 \pi$ for the
commutative case, can represent the characteristic of the
thermodynamical temperature of the smeared systems. The details of
this definition are beyond the scope of this paper.

5. {\it The origin $r=0$ is also smeared by a ``new'' horizon for
$\theta c_{\phi} >0$}: There is a third solution
of $\hat{N}^2=0$ at
\begin{\eq}
\label{IIH} \hat{r}_{--}= l \theta c_{\phi}
\end{\eq}
when $\theta c_{\phi} >0$. This provides a new ``apparent
horizon''\footnote{Similar situation also occurs in AdS black holes
with self-interacting scalar hairs \ci{Mart:06}.}, inside the inner
horizon, encircling the origin. The interior region of
$\hat{r}_{--}$ has two-times as in the region between $\tr$ and
$\hr$.

In the pure $AdS_3$ solutions or the {\it negative mass} black
holes, generally, which can be obtained by considering $r_{\pm}^2
\ra -r_{\pm}^2 $ in the black hole solution (\ref{BTZ}), there is no
event horizon and there appears a naked conical singularity at the
origin. The conical singularity may be smeared by this new horizon
as the results of the spatial noncommutativity

It is also noted that
the appearance of the new horizon near the origin depends on the
sign of $\theta c_{\phi}$, which is analogous to the sign dependence
on the existence of soliton solutions, ``fluxons'' in the field
theories \ci{Doug:01}.

\section*{Acknowledgments}

We would like to thank Hyun Seok Yang for helpful discussions. This
work was supported by the Korea Research Foundation Grant funded by
Korea Government(MOEHRD) (KRF-2007-359-C00011;M.-I.P.)
(KRF-2005-075-C00009;H.-C.K.) and in part by the Korea Science and
Engineering Foundation Grant No. R01-2004-000-10526-0 and through
the Center for Quantum Spacetime (CQUeST) of Sogang University with
grant number R11-2005-021.

\newcommand{\J}[4]{#1 {\bf #2} #3 (#4)}
\newcommand{\andJ}[3]{{\bf #1} (#2) #3}
\newcommand{\AP}{Ann. Phys. (N.Y.)}
\newcommand{\MPL}{Mod. Phys. Lett.}
\newcommand{\NP}{Nucl. Phys.}
\newcommand{\PL}{Phys. Lett.}
\newcommand{\PR}{Phys. Rev. D}
\newcommand{\PRL}{Phys. Rev. Lett.}
\newcommand{\PTP}{Prog. Theor. Phys.}
\newcommand{\hep}[1]{ hep-th/{#1}}
\newcommand{\hepp}[1]{ hep-ph/{#1}}
\newcommand{\hepg}[1]{ gr-qc/{#1}}
\newcommand{\bi}{ \bibitem}

\end{document}